# Optical properties of Ge-oxygen defect center embedded in silica films

F. Messina[1], S. Agnello[1], R. Boscaino[1], M. Cannas*[1], S. Grandi[2], E. Quartarone[2]

[1] Dipartimento di Scienze Fisiche ed Astronomiche dell'Università di Palermo

Via Archirafi 36, I-90123 Palermo, Italy

[2] Dipartimento di Chimica Fisica, Università di Pavia, Via Taramelli 16, I-27100 Pavia, Italy

**ABSTRACT**

The photo-luminescence features of Ge-oxygen defect centers in a 100nm thick Ge-doped silica film on a pure silica substrate were investigated by looking at the emission spectra and time decay detected under synchrotron radiation excitation in the 10-300 K temperature range. This center exhibits two luminescence bands centered at 4.3eV and 3.2eV associated with its de-excitation from singlet ($S_1$) and triplet ($T_1$) states, respectively, that are linked by an intersystem crossing process. The comparison with results obtained from a bulk Ge-doped silica sample evidences that the efficiency of the intersystem crossing rate depends on the properties of the matrix embedding the Ge-oxygen defect centers, being more effective in the film than in the bulk counterpart.

Corresponding author: phone +39 0916235298; fax +39 0916262461, email cannas@fisica.unipa.ir

# 1 INTRODUCTION

Several spectroscopic features of point defects probe the properties of the matrix surrounding them. This is particularly true for those characteristics, such as the non-radiative relaxations, which are activated by the coupling of the optically active center with the environment. Moreover, when point defects are embedded in amorphous media such as silica ($a$-SiO$_2$), they experience the characteristic site-to-site nonequivalence of the solid, this resulting in a inhomogeneous distribution of the observed spectral features [1-3].

Previous works have pointed out that the twofold coordinated Ge (=Ge••), a variety of oxygen deficient Ge centers (GeODC) in bulk SiO$_2$:GeO$_2$ silica, is a model system to study the influence of the amorphous matrix on the optical properties of a defect [3-6]. This center features an absorption band peaked at 5.1eV exciting two photoluminescence (PL) emissions at 4.3eV and 3.2eV, due to de-excitation from excited singlet ($S_1$) and triplet ($T_1$) states [ ], respectively. Moreover, its PL properties agree with a inhomogeneous distribution of the phonon assisted intersystem crossing (ISC) rate, linking $S_1$ and $T_1$, the main experimental evidences deriving from the non-Arrhenius behavior of ISC with temperature and the non-exponential decay of the singlet related emission at 4.3 eV [ ].

This peculiar property of ISC to be influenced by the disordered matrix can be exploited to compare the optical properties of twofold coordinated Ge in bulk silica with its counterpart embedded in different environments such as a thin film, so as to indirectly infer information about the properties of the surrounding. To this aim, in this work we report an experimental study of the PL activity of GeODC centers in a thin Ge-doped SiO$_2$ film sample under pulsed synchrotron radiation excitation.

## 2 EXPERIMENTAL

Experiments described here were carried out on a $SiO_2$:$GeO_2$ film sample (hereafter referred to as "F"), 100nm thick, produced by RF Magnetron Sputtering on a commercial synthetic wet silica substrate. The deposition used a $SiO_2$:$GeO_2$(1% mol) sample prepared with a sol-gel technique as the target, and was performed in a low pressure ($6 \times 10^{-3}$mbar) Ar atmosphere while the substrate was kept at 200°C. We point out that the substrate is optically inactive in the investigated excitation region. For comparison we report also data acquired on a sol-gel $SiO_2$:$GeO_2$ bulk sample (hereafter referred to as "B") 1mm thick.

PL measurements were performed at the I-beamline of SUPERLUMI station at DESY, Hamburg, under 130ps pulse width synchrotron radiation excitation in the UV range with resolution of 0.3 nm. The emitted light was acquired by a liquid nitrogen cooled charge-coupled-device camera for PL spectra and by a photomultiplier (Hamamatsu R2059 model) for lifetime measurements. The emission bandwidth was set 10nm, the spectra were corrected for the spectral response and dispersion of the detection system. Time decay measurements were carried out with a resolution of 0.02 ns. Temperature was varied from 8K to 300K by a Helium-based continuous flow cryostat.

## 3 RESULTS

Fig. 1 plots the evolution of the PL bands peaked at 4.3eV and 3.2eV as detected in the F sample at different temperatures, from 8K to 300K, under excitation at 5.20 eV (corresponding to the maximum in the UV region of the excitation spectrum of the defect). The two bands show an opposite variation with temperature, the PL at 4.3eV decreases by about one order of magnitude whereas the PL at 3.2eV, well observed even at low temperature, increases by a factor ~3. For comparison, we report the emission spectra in the F and B samples detected at 8K, both signals being normalized to the maximum of the 4.3eV band. We note that in the film both the 3.2eV and the 4.3eV emissions are visible, whereas in the bulk sample the 3.2eV emission is almost absent at this temperature.

The different efficiency of triplet emission for the center in bulk and film silica can be quantitatively estimated by the ratio $\eta$ between the intensities of the PL bands at 3.2 eV and 4.3 eV, $\eta=I(3.2eV)/I(4.3eV)$, as a function of temperature T; the corresponding Arrhenius plots of $\ln(\eta)$ with 1/T are shown in Fig. 2. In the F sample $\eta$ increases with temperature by a factor ~20 over the investigated range, moreover it deviates from the Arrhenius behavior as the curve slope decreases on lowering T, and $\eta$ tends to a constant value at T<50K. Also in the B sample $\eta$ does not follow an Arrhenius law and it is smaller than in F sample over all the investigated temperature range, in which its variation is a factor ~50.

Our study of the luminescence properties of GeODC in film silica is completed by measuring the transient kinetics of the 4.3eV PL in the F sample as a function of temperature, the curves being plotted in Fig. 3(a). The time decay assumes a non-exponential character in the overall temperature range and the lifetime $\tau$, measured as the time in which the emission intensity is reduced by a 1/e factor from the maximum, progressively decreases with T from $\tau^F(8K)=4.3\pm0.2$ ns down to $\tau^F(300K)=2.9\pm0.3$ ns. The differences with the bulk counterpart are evidenced in Fig. 3(b): the time decay of the emission detected at 4.3eV at 8K for the B sample is a single exponential curve from which we estimate the lifetime $\tau^B(8K)=8.65\pm0.05$ ns, ~2 times longer than that measured in the F sample at the same temperature.

## 4 Discussion

The above reported findings give a description of the optical features associated with the Ge-ODC in film, also on the basis of their comparison with the same defect in bulk silica. In fact, the two PL bands at 3.2 eV and 4.3 eV observed in the F sample under UV excitation show close spectral similarities with their counterpart in the B sample. Hence, a common energetic level scheme (Fig.

4) associated with the Ge-ODC accounts for the PL features arising from both systems. It consists of a ground singlet $S_0$ and the excited $S_1$ and $T_1$ states. The radiative decay channels from $S_1$ and $T_1$ are accounted for by the rates $K_S$ and $K_T$ respectively, whereas the ISC process linking $S_1$ and $T_1$ is characterized by $K_{ISC}$. Other non radiative channels can be neglected [].

The differences between Film- and Bulk- GeODC can be accounted for by the ISC process. In agreement with the level scheme, the parameter $\eta$ equals the ratio between the ISC rate ($K_{ISC}$) and the pure radiative decay rate from $S_1$ ($K_s$): $\eta=K_{ISC}/K_s$ [4, 5]. Deviations from the Arrhenius dependence from temperature evidenced in the bulk sample have been interpreted as due to inhomogeneity inherent in the ISC process []; hence, we infer that similar effects influence also the ISC for the center in the film environment. Quantitatively, as shown in Fig. 2, $\eta$ is higher in the film than in the bulk material in all the investigated temperature range, this meaning that the ISC process is more efficient for the center embedded in the film matrix: $K^F_{ISC}(T)>K^B_{ISC}(T)$. Moreover, we point out that $\eta$ has a different temperature-dependence in the two samples, since its variation from 8K to 300K is much higher in the B sample. Finally, assuming the Arrhenius law, exp(-E/KT), the slope of the $\eta(T)$ curve measures an 'effective' T-dependent activation energy, which is different for the two samples. For instance, at 300K we find $(140\pm10)\times10^{-3}$eV and $(80\pm10)\times10^{-3}$eV respectively in sample B and F.

Further information on the ISC process comes from the analysis of the decay kinetics from the excited singlet state. Indeed, in agreement with the level scheme, the lifetime is conditioned as well by the ISC, since $\tau=(K_{ISC}+K_S)^{-1}$. The evidence of $\tau$ being lower in film than in bulk sample at all temperatures is consistent with a higher efficiency of the ISC process, which in the film remains active and competitive with the radiative channel even at cryogenic temperatures. At variance, in the bulk centers, $K^B_{ISC}(8K)<<K_S$, so that the measured lifetime is purely radiative. If we assume $K_S$ to be the same for the GeODC in F and B, the comparison in Fig. 3(B) between $\tau^F(8K)$ and $\tau^B(8K)$ in the two samples permits to derive $K^F_{ISC}(8K)\approx K_S$.

Moreover, it is known that the deviation of the 4.3eV time decay curve from a single exponential is due to the contribution to the decay of the inhomogeneously distributed $K_{ISC}$. For GeODCs in a bulk environment, a progressive stretching of the decay kinetics is observed above ~100K, [] where the ISC process becomes competitive with the radiative decay. At variance, in the F sample the kinetics is not single-exponential even at 8K.

It is worth to note that the influence of the environment on the ISC process had already been pointed out in previous works on surface GeODC in activated Ge-doped silica samples and was related to differences in the O-Ge-O angle from bulk to surface. Comparison with present results corroborate this finding, though the microscopic parameter determining this sensitivity is not yet clarified.

## 5  Conclusions

The results here reported show that the spectroscopic properties of GeODC centers depend on the host silica matrix. In particular, we found that the phonon assisted ISC results to be more effective when the defect is embedded in a thin film environment than in bulk Ge-doped silica. In both cases, $k_{ISC}$ reflects the inhomogeneity of the amorphous host matrix, so being a probe of its structural disorder.

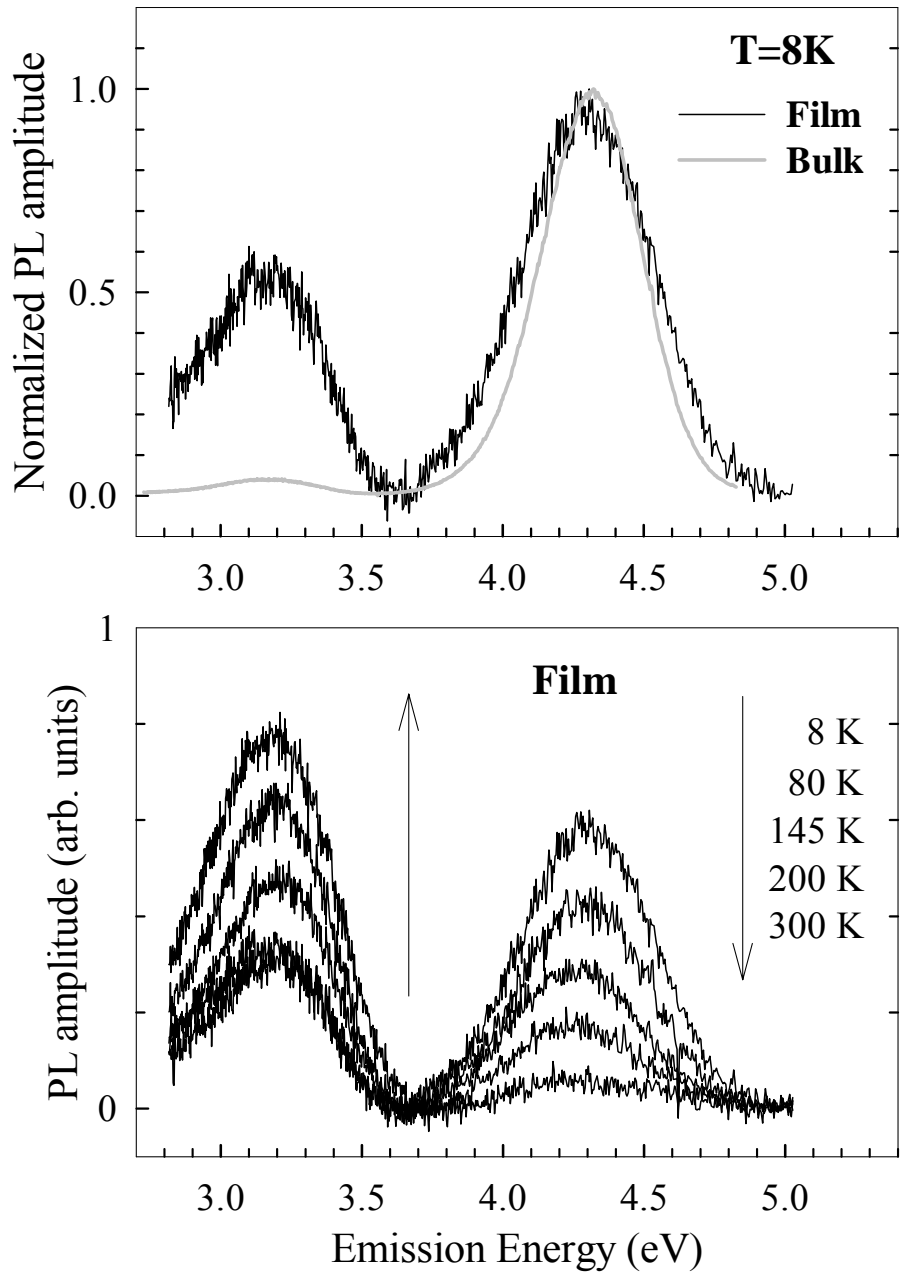

Fig. 1

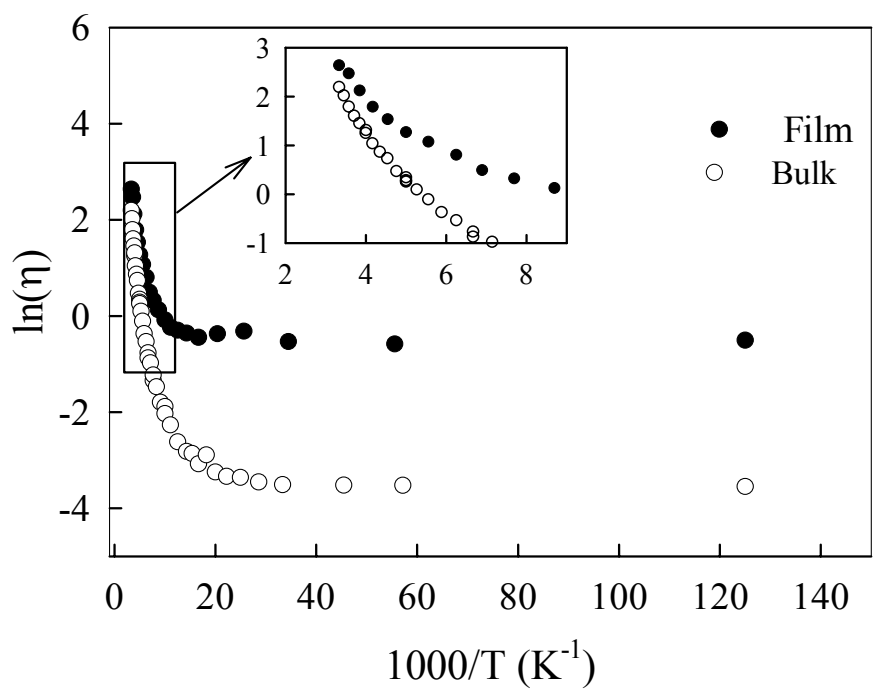

Fig. 2

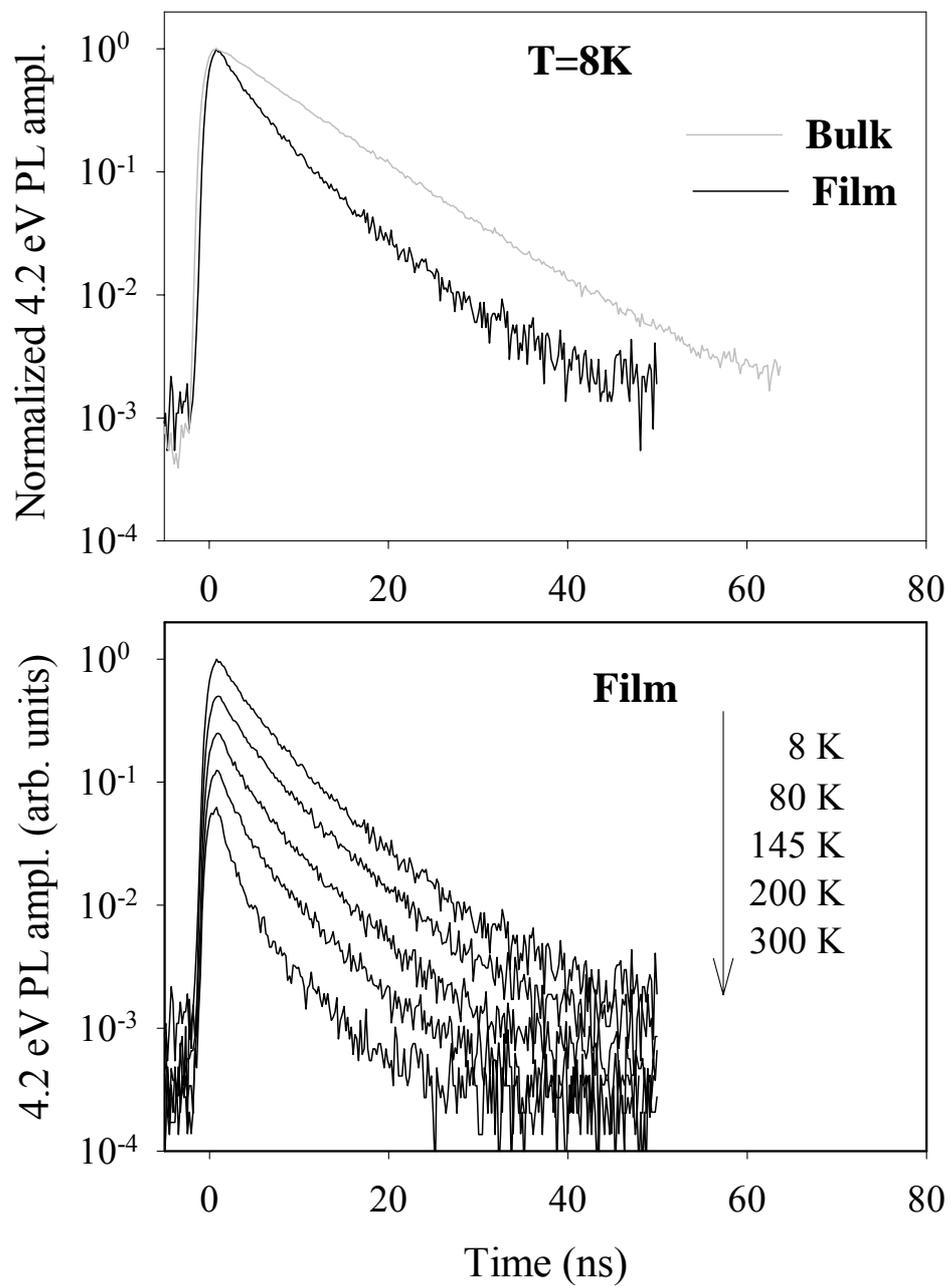

Figure 3